\documentclass[10pt,letterpaper]{article}
\pdfoutput=1

\usepackage{jcappub}
\usepackage{amsmath,amssymb,color}
\usepackage{enumerate,xstring}
\usepackage{jcappub}
\usepackage{amsmath}
\usepackage{amsfonts}
\usepackage{amssymb}
\usepackage{graphicx}
\usepackage{epsfig}
\usepackage{color}
\usepackage{multirow}
\usepackage{graphicx}
\usepackage{bigints}
\usepackage{todonotes,bm,etoolbox}
\usepackage{calligra}
\usepackage{hyperref}
\usepackage{makecell}
\usepackage{tabularx}
\usepackage{booktabs}
\usepackage{subfigure}
\usepackage{soul}

\usepackage{tikz}

\def\ba#1\ea{\begin{align}#1\end{align}}
\def\bea{\begin{eqnarray}}
\def\eea{\end{eqnarray}}
\def\be{\begin{equation}}
\def\ee{\end{equation}}
\def\d{\delta}

\def\({\left(}
\def\){\right)}
\def\[{\left[}
\def\]{\right]}

\def\<{\left\langle}
\def\>{\right\rangle}

\def\comment#1{}

\def\d{{\delta}}
\def\eps{\epsilon}

\renewcommand{\v}[1]{\bm{#1}}

\def\vk{\v{k}}

\newcommand{\perm}[1]{ \expandafter\ifstrempty\expandafter{#1} {\mbox{perm.}} {\mbox{$#1$ perm.}} }

\DeclareMathOperator{\Cov}{\rm \bf Cov}

\newcommand{\fnl}{f_\textnormal{\textsc{nl}}}

\newcommand{\bphi}{b_\phi}
\newcommand{\bphia}{\langle b_\phi \rangle}
\newcommand{\minv}{\mathcal{M}(k)^{-1}}

\newcommand{\bfac}{|b_1^B \bphi^A - b_1^A\bphi^B|}

\def\pathtofigs{./}

\definecolor{RedWine}{rgb}{0.743,0,0}
\definecolor{RoyalBlue}{rgb}{0.25,.41,.88}
\definecolor{ForestGreen}{rgb}{.13,.54,.13}
\definecolor{Goldenrod}{rgb}{.85,.65,.13}

\newcommand{\bq}{\begin{eqnarray}}
\newcommand{\eq}{\end{eqnarray}}

\title{Towards optimal and robust $\fnl$ constraints with multi-tracer analyses}

\author[a,b]{Alexandre Barreira}
\author[c]{and Elisabeth Krause}

\affiliation[a]{\small Excellence Cluster ORIGINS, Boltzmannstra\ss e 2, 85748 Garching, Germany}
\affiliation[b]{\small Ludwig-Maximilians-Universit\"at, Schellingstra\ss e 4, 80799 M\"unchen, Germany}
\affiliation[c]{\small Department of Astronomy and Steward Observatory, University of Arizona, 933 North Cherry Ave, Tucson, Arizona 85721, USA}

\emailAdd{alex.barreira@origins-cluster.de, krausee@arizona.edu}

\date{\today}

\abstract{We discuss the potential of the multi-tracer technique to improve observational constraints of the local primordial non-Gaussianity (PNG) parameter $\fnl$ from the galaxy power spectrum. For two galaxy samples $A$ and $B$, the constraining power is $\propto \bfac$, where $b_1$ and $\bphi$ are the linear and PNG galaxy bias parameters. We show this allows for significantly improved constraints compared to the traditional expectation $\propto |b_1^A - b_1^B|$ based on naive universality-like relations where $\bphi \propto b_1$. Using IllustrisTNG galaxy simulation data, we find that different equal galaxy number splits of the full sample lead to different $\bfac$, and thus have different constraining power. Of all of the strategies explored, splitting by $g-r$ color is the most promising, more than doubling the significance of detecting $\fnl\bphi \neq 0$. Importantly, since these are constraints on $\fnl\bphi$ and not $\fnl$, they do not require priors on the $\bphi(b_1)$ relation. For direct constraints on $\fnl$, we show that multi-tracer constraints can be significantly more robust than single-tracer to $\bphi$ misspecifications and uncertainties; this relaxes the precision and accuracy requirements for $\bphi$ priors. Our results present new opportunities to improve our chances to detect and robustly constrain $\fnl$, and strongly motivate galaxy formation simulation campaigns to calibrate the $\bphi(b_1)$ relation.}

\begin{document}

\maketitle

\section{Introduction}
\label{sec:intro}

Constraining the local primordial non-Gaussianity (PNG) parameter $\fnl$ \cite{2001PhRvD..63f3002K} is one of the most powerful ways to learn about the physics of inflation. A detection of $\fnl \neq 0$ could rule out single-field inflation \cite{maldacena:2003, 2004JCAP...10..006C, 2011JCAP...11..038C, Tanaka:2011aj, conformalfermi}, and constraining $|\fnl| \lesssim 1$ can be used to test many other classes of models \cite{2014arXiv1412.4872D, 2019Galax...7...71B, 2022arXiv220308128A}. The current tightest bound is $\fnl = -0.9 \pm 5.1\ (1\sigma)$ from cosmic microwave background data \cite{2020A&A...641A...9P}, and the next improvements over it are expected to come from galaxy clustering analyses \cite{2014arXiv1412.4872D, 2019Galax...7...71B, 2022arXiv220308128A}.

\vspace{1mm}
In local PNG cosmologies, the linear galaxy density contrast contains two terms given by \cite{slosar/etal:2008, mcdonald:2008, giannantonio/porciani:2010, 2011JCAP...04..006B, assassi/baumann/schmidt}
\be
\label{eq:biasexp}
\d_g(\vk,z) \supset b_1(z) \d_m(\vk,z) + \bphi(z) \fnl\phi(\vk),
\ee
where $\vk$ is the wavevector, $\d_m$ is the matter density contrast, $\phi$ is the primordial potential in the matter era, and $b_1$ and $\bphi$ are galaxy bias parameters that describe the response of galaxy formation to large-scale $\delta_m$ and $\phi$ perturbations, respectively \cite{biasreview}. The contribution from $\fnl$ in Eq.~(\ref{eq:biasexp}) is called the {\it scale-dependent bias} effect \cite{dalal/etal:2008}, and is how galaxy data are most directly sensitive to $\fnl$. Some of the latest reported constraints using galaxy data include $\fnl = -12 \pm 21\ (1\sigma)$ using eBOSS quasars \cite{2021arXiv210613725M} (see also Ref.~\cite{2019JCAP...09..010C}), and $\fnl = -30 \pm 29\ (1\sigma)$ \cite{2022arXiv220111518D} and $\fnl = -33 \pm 28\ (1\sigma)$ \cite{2022PhRvD.106d3506C} using galaxies in the BOSS survey. Note however that these constraints require knowledge of the $\bphi(b_1)$ relation to break the $\fnl\bphi$ degeneracy in Eq.~(\ref{eq:biasexp}). This relation is still uncertain \cite{2010JCAP...07..013R, 2017MNRAS.468.3277B, 2020JCAP...12..013B, Voivodic:2020bec, 2022JCAP...01..033B, 2022JCAP...04..057B, 2023JCAP...01..023L}, and so these bounds on $\fnl$ are currently subject to significant model systematic errors \cite{2020JCAP...12..031B, 2022JCAP...11..013B}; independently of $\bphi$, only the parameter combination $\fnl\bphi$ can be robustly constrained, and not $\fnl$.

\vspace{1mm}
The {\it multi-tracer} technique \cite{2009PhRvL.102b1302S} is a tool to increase the statistical constraining power of a given galaxy sample. The idea is to split the galaxy sample into sub-samples occupying the same volume. Since the sub-samples formed from the same initial density field, at sufficiently high number density, the relative clustering amplitude of the sub-samples can be measured effectively noise-free and can be used to constrain $\fnl$ with very high precision (see Refs.~\cite{2022JCAP...04..021M, 2023arXiv230605474M} for recent studies of the impact of multi-tracer analyses to constrain other cosmological parameters). In doing so, relations like the {\it universality relation} $\bphi(b_1) = 2\delta_c(b_1-1)$, $\delta_c = 1.686$, where $b_1$ and $\bphi$ vary proportionally to each other are traditionally assumed in $\fnl$ constraints, which for multi-tracer analyses leads to the constraining power being maximized for large differences in $b_1$ of the sub-samples. However, a number of recent works using galaxy formation simulations \cite{2020JCAP...12..013B, Voivodic:2020bec, 2022JCAP...01..033B, 2022JCAP...04..057B} have shown that these universality-based relations are likely not good descriptions of real-life galaxies, making it interesting to revisit this expectation.

\vspace{1mm}
In this paper we use the possibility that different galaxy types may have different $\bphi(b_1)$ relations to investigate optimal galaxy selection strategies for $\fnl$ multi-tracer constraints.  For splits into two sub-samples $A$ and $B$, the constraining power is $\propto \bfac$ \cite{2009PhRvL.102b1302S}, and not simply $\propto |b_1^A - b_1^B|$ as when naively assuming the universality relation. That is, the constraining power is the greatest for multi-tracer splits in which $\bphi$ and $b_1$ vary inversely proportional to each other, which is orthogonal to the case enforced by the universality relation. Interestingly, using galaxy formation simulations, we are able to identify multi-tracer splits, e.g.~based on galaxy color, that can have large values of $\bfac$ and thus exploit much better the power of the multi-tracer technique to detect $\fnl \neq 0$.

\vspace{1mm}
We also investigate the impact that $\bphi$ priors have in multi-tracer analyses. Compared to single-tracer, we find that multi-tracer analyses are more robust to $\fnl$ biases due to $\bphi$ misspecifications, as well as degradation in $\fnl$ constraining power due to $\bphi$ uncertainties. This motivates further investing in the multi-tracer technique in $\fnl$ constraints as a way to relax the accuracy and precision requirements on $\bphi$ priors calibrated from galaxy formation simulations.

\vspace{1mm}
The rest of this paper is as follows. In Sec.~\ref{sec:info} we study the impact that $b_1$ and $\bphi$ have on the $\fnl$ Fisher information in single- and multi-tracer cases. In Sec.~\ref{sec:sims} we show the improvements in $\fnl$ constraints from multi-tracer analyses using galaxies from hydrodynamical simulations. We investigate the robustness of single- and multi-tracer constraints to priors on $\bphi$ in Sec.~\ref{sec:priors}, and conclude in Sec.~\ref{sec:summary}.

\vspace{1mm}
For all numerical results we assume a fictitious galaxy survey at $z=1$ with volume $V_S = 100\ {\rm Gpc^3}/h^3$, and consider multi-tracer splits into two sub-samples $A, B$ with equal number density, i.e.~$\bar{n}_g^A = \bar{n}_g^B = \bar{n}_g / 2$. We assume the spectra is measured from $k_{\rm min}=\pi/V_S^{1/3}$ up to $k_{\rm max} = 0.2 h/{\rm Mpc}$ in 40 log-spaced $k$ bins. We adopt the following cosmological parameters throughout: $\Omega_m = 0.3089$, $\Omega_b = 0.0486$, $\Omega_{\Lambda} = 0.6911$, $H_0 = 100 h\ {\rm km/s/Mpc}$ with $h = 0.6774$, $n_s = 0.967$, $\sigma_8 = 0.816$ and $\fnl = 5$. We note our main conclusions do not change under reasonable modifications to any of these values. We use the  {\sc CAMB} code \cite{camb} to calculate power spectra and transfer functions.

\section{Fisher information analysis}
\label{sec:info}

In this section, we discuss the impact that different values of the galaxy bias parameters $b_1$ and $\bphi$ have on the $\fnl$ Fisher information.

\subsection{Single-tracer case}
\label{sec:error_st}

The large-scale power spectrum of a single galaxy sample can be written in real space as
\bq\label{eq:model_st}
P_{gg}(k,z) &=& \Big[b_1(z) + \bphi(z) \fnl \mathcal{M}(k,z)^{-1}\Big]^2 P_{mm}(k,z)+ P_{\eps},
\eq
where $P_{mm}$ is the linear matter power spectrum and $P_{\eps} = 1/\bar{n}_g$ is the shot noise (which we assume Poissonian). The quantity $\mathcal{M}$ is defined as $\delta_m(\vk, z) = \mathcal{M}(k,z)\phi(\vk)$, and given by $\mathcal{M}(k,z) = (2/3)k^2T_m(k) D_{\rm md}(z)/(\Omega_m H_0^2)$, where $T_m$ is the matter transfer function and $D_{\rm md}$ is the linear growth factor normalized to $1/(1+z)$ during matter domination. The Fisher information on $\fnl$ is
\bq\label{eq:fisher_info_st_1}
F_{\fnl}(k,z) = \frac{\partial P_{gg}(k,z)}{\partial \fnl} {\rm Cov}^{-1}(k, z) \frac{\partial P_{gg}(k,z)}{\partial \fnl},
\eq
where ${\rm Cov}$ is the data covariance matrix. As we focus on $\fnl$ constraints, which are dominated by the largest scales, we approximate the covariance by the diagonal disconnected component
\bq\label{eq:cov_st_1}
{\rm Cov}(k,z) = \frac{2(2\pi)^3}{V_SV_{k}} P_{gg}(k,z)^2,
\eq
where $V_S$ is the survey volume, $V_k = 4\pi k^2 \Delta k$, and $\Delta k$ is the wavenumber bin size; from hereon we suppress the redshift dependence to ease the notation.

The error bar on $\fnl$ is given by $\sigma_{\fnl} = 1/\sqrt{F_{\fnl}}$, which reads as 
\bq\label{eq:fisher_error_st_1}
\sigma_{\fnl}(k) = \sqrt{\frac{2(2\pi)^3}{V_SV_{k}}} \frac{b_1 + \bphi \fnl \minv}{2\bphi\minv} \left(1 + X(k)\right),
\eq
where we introduced $X(k) = P_\eps/(P_{gg}(k) - P_{\eps})$ to quantify the impact of shot noise. Concerning the bias parameters, this expression tells us that the constraining power on $\fnl$ increases with larger values of $\bphi$. This is for two reasons. First, at fixed number density $\bar{n}_g$, increasing $\bphi$ lowers $X$, which decreases $\sigma_{\fnl}$. Second, the precision on $\fnl$ increases also if $\bphi \gg b_1$ as the Gaussian contribution to the power spectrum $\propto b_1^2$ does not contain information on $\fnl$, but still contributes to the total variance in Eq.~(\ref{eq:cov_st_1}); i.e., the limit $\bphi \gg b_1$ maximizes the signal-to-noise on $\fnl$. This is the idea explored in Ref.~\cite{2018PhRvL.121j1301C} with {\it zero-bias} tracers in which one attempts to select a galaxy sample with $b_1 = 0$ that suffers only from a cosmic variance error given by
\bq\label{eq:fisher_error_st_2}
\sigma_{\fnl}(k)   = \sqrt{\frac{2(2\pi)^3}{V_SV_{k}}} \frac{\fnl}{2} \left(1 + X(k)\right).
\eq

The takeaway for single-tracer analyses is thus that, at fixed number density $\bar{n}_g$, {\it the optimal $\fnl$ constraints are obtained with samples that have the largest values of $\bphi$ and the lowest values of $b_1$}.

\subsection{Multi-tracer case}
\label{sec:error_mt}

We consider multi-tracer analyses in which a full galaxy sample is split into two equal-sized sub-samples labeled as $A$ and $B$. The data vector is now composed of the power spectra and cross-spectra of the two samples ${\bf D}(k) = \Big\{P_{gg}^{A}(k), P_{gg}^{A\times B}(k), P_{gg}^{B}(k)\Big\}$, and the Fisher information is 
\bq\label{eq:fisher_info_mt_1}
F_{\fnl}(k) &=& \frac{\partial {\bf D}^t(k)}{\partial \fnl} \cdot {\bf Cov}^{-1}(k) \cdot  \frac{\partial {\bf D}(k)}{\partial \fnl},
\eq
where ${\bf Cov}(k)$ is the covariance of the data vector ${\bf D}$ (cf.~Eq.~(\ref{eq:cov_app_1})). In the cosmic variance limit $X^{A}, X^{B} \ll 1$, the result is given by (we outline a few intermediate steps of the derivation in App.~\ref{app:derivation})
\bq\label{eq:fisher_info_mt_2}
F_{\fnl} = 2\frac{V_S V_k}{2(2\pi)^3} \frac{1}{X^B + X^A} \left(\frac{\left(b_1^B \bphi^A - b_1^A\bphi^B\right)\minv}{R \left(b_1^B + \fnl\bphi^B\minv\right)^2}\right)^2 + \mathcal{O}\left(X^A, X^B\right),
\eq
where $R = \big(P_{gg}^B - P_\eps^B\big)/\big(P_{gg}^A - P_\eps^A\big)$, and the superscripts label the various quantities for samples $A$ and $B$; for example $X^{A}=P_{\eps}^{A}/\left(P_{gg}^{A}-P_{\eps}^{A}\right)$. One of the important impacts of galaxy bias comes through the value of $\bfac$ in the numerator of Eq.~(\ref{eq:fisher_info_mt_2}); this is as previously derived in Ref.~\cite{2009PhRvL.102b1302S} (see their Eq.~(10)). Effectively all works in the literature assume the universality relation $\bphi(b_1) = 2\delta_c(b_1-1)$ (or some variant of it with $\bphi \propto b_1$), from which it follows that $F_{\fnl}$ increases with increasing $|b_1^A - b_1^B|$. This is why multi-tracer analyses were traditionally described as having greater constraining power when the sub-samples have very different $b_1$. In this paper, however, we highlight that the situation is actually richer than what universality naively suggests: {\it in the limit of low shot noise, the precision on $\fnl$ increases with increasing $|b_1^B \bphi^A - b_1^A\bphi^B|$, and not just $|b_1^A - b_1^B|$.}\footnote{This can be seen also as the $\fnl$ precision being $\propto {\rm sin}\alpha$, with $\alpha$ the angle between the vectors $(b_1^A, \bphi^A)$ and $(b_1^B, \bphi^B)$.}

\vspace{1mm}

This is interesting as a number of studies \cite{2010JCAP...07..013R, 2017MNRAS.468.3277B, 2020JCAP...12..013B, Voivodic:2020bec, 2022JCAP...01..033B, 2022JCAP...04..057B, 2023JCAP...01..023L} have found using simulations that the $\bphi(b_1)$ relation is a very sensitive function of the details of halo and galaxy formation physics. Thus, provided we are able to understand the $\bphi(b_1)$ relation of real galaxies sufficiently well, this gives us a chance to optimize our galaxy selection strategies by exploring cases that maximize the parameter combination $\bfac$. Indeed, as we will see throughout, this can lead to significant improvements compared to the traditional expectation based on the universality relation.

\subsection{Simple illustration of the impact of galaxy bias}
\label{sec:illustration}

\begin{figure}
\centering
\includegraphics[width=1.0\textwidth]{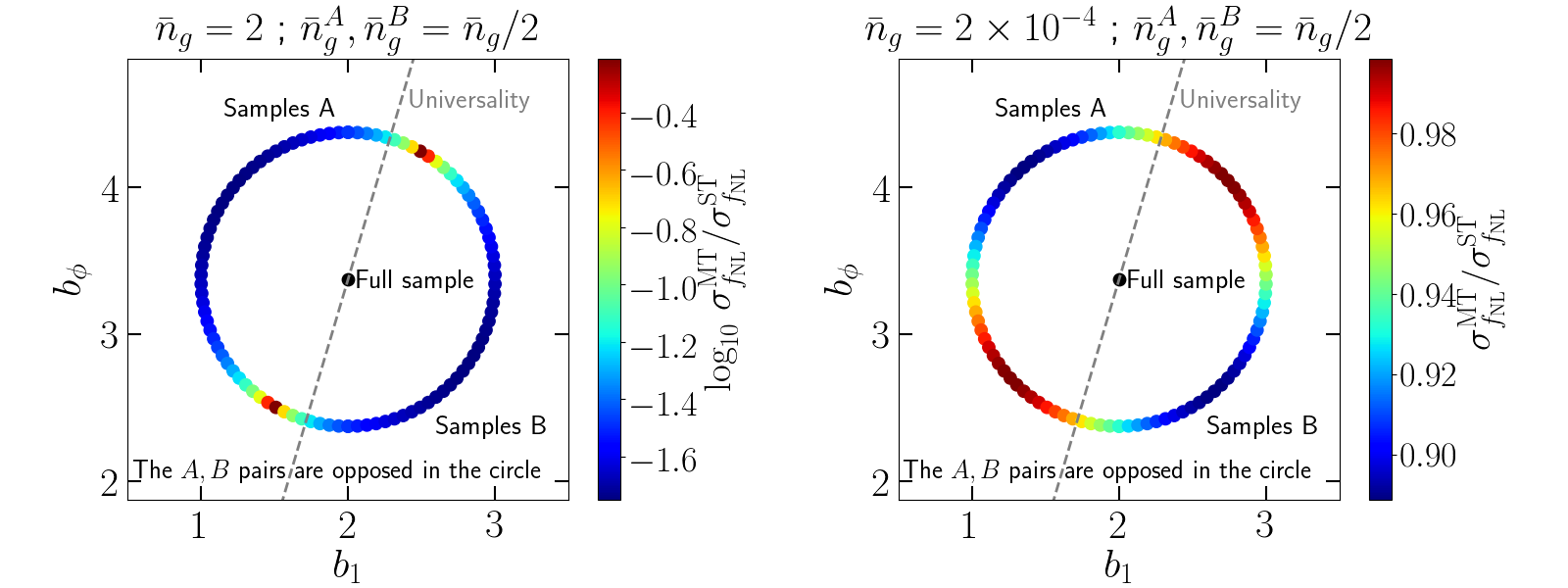}
\caption{Relative improvement in $\fnl$ constraints for different values of $b_1$ and $\bphi$ in multi-tracer (MT) analyses. The pairs of galaxy sub-samples are chosen to be diametrically opposed on a unit radius circle around the full sample values $b_1 = 2$ and $\bphi = 3.37$. Concretely, $b_1^B = b_1 + {\rm cos}(\theta)$, $\bphi^B = \bphi + {\rm sin}(\theta)$ and $b_1^A = b_1 + {\rm cos}(\theta + \pi)$, $\bphi^A = \bphi + {\rm sin}(\theta + \pi)$, with $\theta \in \left[-\pi/2, \pi/2\right]$; this simple parametrization serves only the purpose to illustrate the typical impact of $b_1$ and $\bphi$. The result is for equal number density splits $\bar{n}_g^A = \bar{n}_g^B = \bar{n}_g/2$; the left and right panels are for an unrealistically large and realistic value of $n_g$, as labeled (the $\bar{n}_g$ units are $h^3/{\rm Mpc^3}$). The colors indicate the improvement of the multi-tracer constraints relative to the single-tracer (ST) result using the full galaxy sample. The grey dashed line shows the universality relation.}
\label{fig:circle}
\end{figure}

Figure~\ref{fig:circle} compares the constraining power of different values of $b_1$ and $\bphi$ in multi-tracer $\fnl$ constraints. In this toy illustration we place the sub-samples $A$ and $B$ on a unit circle centered around the $b_1$ and $\bphi$ values of the full sample, with each $A,B$ sample pair linking diametrically opposed points. The values of the full sample are $b_1 = 2$ and $\bphi = 2\delta_c(b_1-1) = 3.37$. The left panel shows the result for an unrealistically large value of the number density $\bar{n}_g = 2\ {h^3/{\rm Mpc^3}}$ to enforce the cosmic variance limited regime, and the right panel shows the result for a realistic value of $\bar{n}_g = 2\times 10^{-4}\ {h^3/{\rm Mpc^3}}$. 

In the cosmic variance limit on the left of Fig.~\ref{fig:circle}, the improvement is largest (dark blue) when the $b_1$ and $\bphi$ values of the sub-samples vary inversely proportional to each other. This is as expected from Eq.~(\ref{eq:fisher_info_mt_2}) since this is what maximizes the $\bfac$ factor. Conversely, when $\bfac \approx 0$, the improvements from the multi-tracer technique become much less pronounced (cf.~red and yellow points on the left). Further, the grey dashed line shows the universality relation, which is not aligned with the directions in the $b_1 - \bphi$ plane that give the best constraints.

For the realistic number density scenario on the right of Fig.~\ref{fig:circle}, the $\mathcal{O}\left(X^A, X^B\right)$ terms in Eq.~(\ref{eq:fisher_info_mt_2}) spoil the exact scaling with $\bfac$, but the qualitative picture remains the same: there is still a correlation between the improvement of the constraints and the value of $\bfac$, which happens for directions in the $b_1 - \bphi$ that are orthogonal to the universality direction. Note that the quantitative improvements in Fig.~\ref{fig:circle} are specific to our illustrative choice of bias parameters; below we discuss results for more realistic values of $b_1$ and $b_\phi$.

\section{Multi-tracer examples from galaxy formation simulations}
\label{sec:sims}

In the last section we have identified that multi-tracer analyses that maximize the value of $\bfac$ are those that return optimal constraints on $\fnl$. What we wish to do now using galaxy formation simulations is to identify which galaxy selection strategies can lead to larger values of $\bfac$, and are thus optimal to constrain $\fnl$ using the multi-tracer technique.

\subsection{Simulated galaxy samples}
\label{sec:sims_specs}

\begin{table*}
\centering
\begin{tabular}{lcccccccccccccc}
\toprule
&  & $M_t$ & $M_*$ & $g-r$ & $\dot{M}_{\rm BH}$ & $n_{g,8}$ & $c_{200}$ (halos) \\
\midrule
\midrule
& $\left(b_1^A, b_1^B\right)$ &  $\left(3.0, 2.5\right)$ & $\left(3.0, 2.5\right)$ & $\left(2.8, 2.7\right)$ & $\left(3.2, 2.2\right)$ & $\left(5.4, 0.1\right)$ & $\left(2.4, 3.0\right)$  \\
\midrule
& $\left(\bphi^A, \bphi^B\right)$ &  $\left(5.0, 3.9\right)$ & $\left(7.3, 1.6\right)$ & $\left(9.1, -0.2\right)$ & $\left(4.3, 4.7\right)$ & $\left(9.1, -0.1\right)$ & $\left(6.9, 3.6\right)$  \\
\midrule
& $\bfac$ & $0.5$ &  $13.5$ & $25.0$ & $5.5$ & $1.4$ & $11.6$  \\
\midrule
\bottomrule
\end{tabular}
\caption{Numerical values of $b_1$ and $\bphi$ of the simulated galaxy samples used in this work. The columns for $M_t$, $M_*$, $g-r$, $\dot{M}_{\rm BH}$ and $n_{g,8}$ are for splits of a galaxy sample with $b_1 = 2.7$ and $\bphi = 4.5$. The column for $c_{200}$ is for a split of a halo sample with $b_1 = 2.7$ and $\bphi = 5.3$. All samples are at $z=1$ and the full sample has $\bar{n}_g = 2\times10^{-4}\ h^3/{\rm Mpc^3}$. All splits are at equal number density $\bar{n}_g^A = \bar{n}_g^B = \bar{n}_g/2$ using the $50\%$ highest (sample A) and lowest (sample B) values of each property. The last line lists the parameter combination $\bfac$ that controls the constraining power of multi-tracer analyses on $\fnl$ (cf.~Eq.~(\ref{eq:fisher_info_mt_2})).}
\label{tab:sims}
\end{table*}

\begin{figure}
\centering
\includegraphics[width=1.0\textwidth]{\pathtofigs 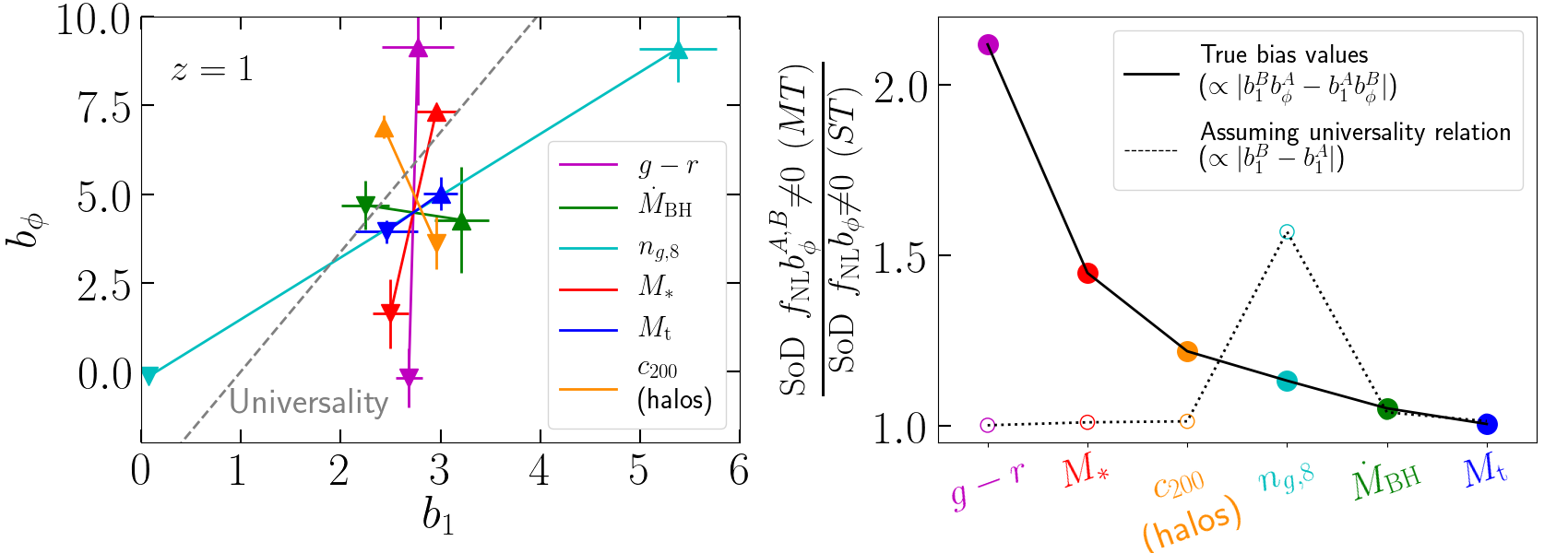}
\caption{Improvements on the significance of detection (SoD) of $\fnl \neq 0$ for several equal-number multi-tracer (MT) splits using halos and galaxies in simulations. The full sample is obtained with a minimum total mass cut that results in $\bar{n}_g = 2\times 10^{-4}\ {h^3}/{\rm Mpc^3}$ at $z=1$. The two sub-samples are constructed by splitting this sample in half according to the values of different galaxy properties, as labeled: total mass $M_t$, stellar mass $M_*$, $g-r$ color, black hole mass accretion rate $\dot{M}_{\rm BH}$, and environmental density $n_{g,8}$. The resulting $b_1$ and $\bphi$ values are marked on the left panel and in Tab.~\ref{tab:sims}. The solid line and filled symbols in the right panel show the improvement on the significance-of-detection (SoD) of $\fnl$ relative to single-tracer (ST) from Fisher forecasts for a survey with $V_S = 100\ {\rm Gpc}^3/h^3$ at $z=1$. The dotted line and open symbols show the same but assuming the universality relation to highlight the gains from taking into account the true bias parameters.}
\label{fig:sims}
\end{figure}

We utilize the $N$-body simulation data of Refs.~\cite{2020JCAP...12..013B, 2022JCAP...01..033B}, concretely a suite of gravity-only and galaxy formation simulations run with the IllustrisTNG galaxy formation model \cite{2018MNRAS.473.4077P}. We refer the reader to Refs.~\cite{2020JCAP...12..013B, 2022JCAP...01..033B} (as well as to Ref.~\cite{2023JCAP...01..023L}) for the details about how the $b_1$ and $\bphi$ parameters are measured using the separate universe approach. 

We will show results for galaxy samples constructed as follows. The full galaxy sample is obtained by applying a minimum total mass cut at $z=1$ that yields a total number density of $\bar{n}_g = 2\times10^{-4}\ h^3/{\rm Mpc^3}$.\footnote{We have repeated the analysis for $z = 0, 0.5, 1, 2, 3$, as well as for higher ($\bar{n}_g = 5\times10^{-4}\ h^3/{\rm Mpc^3}$) and lower ($\bar{n}_g = 1\times10^{-4}\ h^3/{\rm Mpc^3}$) number densities, and found qualitatively consistent results. Analyses with a full sample selected by a minimum stellar mass cut revealed also the same main takeaway points.} This sample is then split into two sub-samples that have the $50\%$ lowest and highest values of different galaxy properties. We consider splits by: (i) total mass $M_t$, (ii) stellar mass $M_*$, (iii) $g-r$ color, (iv) black hole mass accretion rate $\dot{M}_{\rm BH}$ and (v) environmental density $n_{g,8}$, defined as the number density of neighbour galaxies within $8\ {\rm Mpc}/h$, which we estimate using the method of Ref.~\cite{2019MNRAS.483.4501A}. In addition, we consider also a main halo population constructed in the same way, but split into two by their Navarro-Frenk-White $c_{200}$ concentration \cite{2023JCAP...01..023L}. Our halo results are for a gravity-only simulation with $L_{\rm box}=560\ {\rm Mpc}/h$ and $N_p = 1250^3$ tracer particles, while our galaxy results are for simulations with IllustrisTNG with $L_{\rm box}=205\ {\rm Mpc}/h$ and $N_p = 2\times1250^3$. The cosmology of these simulations is the same as that adopted in this paper, except that $\fnl = 0$ (cf.~Sec.~\ref{sec:intro}).

The left panel of Fig.~\ref{fig:sims} shows the $b_1$ and $\bphi$ values of all these samples, as labeled; they are also listed in Tab.~\ref{tab:sims}. The value for the full sample is identified by where all lines intercept (except the orange one for $c_{200}$). The orange line for $c_{200}$ for halos does not intersect the same point because the full halo population has slightly different bias values compared to the full galaxy sample. A remarkable aspect of the result is that different ways to split the same galaxy sample into two equal number density sub-samples can shift the bias parameters in markedly different directions in the $b_1-\bphi$ plane. That is, at least within the astrophysics of the IllustrisTNG model, there is indeed room to explore different galaxy selection strategies to optimize multi-tracer $\fnl$ constraints.

\subsection{Constraints for the simulated galaxy samples}
\label{sec:sims_specs}

The solid line and filled symbols in the right panel of Fig.~\ref{fig:sims} show the ratio of the significance of detection (SoD) of $\fnl \neq 0$ for the various multi-tracer splits and the single-tracer result with the full sample. Concretely, we use a Fisher forecast to constrain the parameter combination $\fnl\bphi$, which lets us assess the SoD without priors on $\bphi$. Note that detecting $\fnl\bphi \neq 0$ implies $\fnl \neq 0$, so it is still possible to learn something about inflation (namely that it was not single-field) without knowing the actual value of $\fnl$. To date, the only examples of constraints placed directly on $\fnl\bphi$ were obtained in Refs.~\cite{2022PhRvD.106d3506C, 2022JCAP...11..013B} using BOSS DR12 galaxies. For single-tracer analyses, the SoD is defined as ${\rm SoD}^{\rm ST} = \langle{\fnl\bphi}\rangle/\sigma_{\fnl\bphi}$, where $\langle{\fnl\bphi}\rangle$ and $\sigma_{\fnl\bphi}$ are the mean of the constraints and the $1\sigma$ error bar, respectively. For multi-tracer samples, the SoD is given by
\bq\label{eq:sodmt}
{\rm SoD}^{\rm MT} = \sqrt{\left(\langle{\fnl\bphi^A}\rangle, \langle\fnl\bphi^B\rangle\right)^t \cdot F_{\fnl\bphi^A,\fnl\bphi^B} \cdot \left(\langle{\fnl\bphi^A}\rangle, \langle\fnl\bphi^B\rangle\right)},
\eq
where $\langle{\fnl\bphi^A}\rangle$ and $\langle{\fnl\bphi^B}\rangle$ are the mean of the constraints and $F_{\fnl\bphi^A,\fnl\bphi^B}$ is their Fisher matrix.

\vspace{1mm}

The result in Fig.~\ref{fig:sims} shows that the choice of multi-tracer strategy can have a significant impact on the SoD of $\fnl \neq 0$. The corresponding values of $\bfac$ are listed in Tab.~\ref{tab:sims}, which display the expected correlation with the improvement of the SoD from Eq.~(\ref{eq:fisher_info_mt_2}). The only exception is the split by $\dot{M}_{\rm BH}$, which performs worse than $n_{g,8}$, despite the larger $\bfac$; this is because the $\bfac$ scaling in Eq.~(\ref{eq:fisher_info_mt_2}) is only {\it exact} in the cosmic variance limit. Of all the multi-tracer cases considered, the split by $g-r$ color performs the best, more than doubling the SoD of $\fnl$ compared to single-tracer.\footnote{Note that our $g-r$ split is not equal to the traditional split into star-forming (blue) and quiescent (red) galaxies.} We emphasize that these improvements can normally be achieved for {\it free} if galaxy imaging data exist, i.e.~they do not require additional spectroscopic observations.

\vspace{1mm}

In order to highlight the differences to the traditional approach in the literature, the dotted line and open symbols show the same, but incorrectly assuming that $\bphi$ is determined from $b_1$ using the universality relation.  The differences are significant, strongly motivating abandoning assuming the universality relation in multi-tracer $\fnl$ constraints. In particular, the universality expectation $\propto |b_1^A-b_1^B|$ misleadingly characterizes the $g-r$ multi-tracer split as very weak, despite it being the best in this work. Conversely, the two $n_{g,8}$ samples have the largest difference in $b_1$, and so are misleadingly classified as being the best multi-tracer split, despite it being actually one of the worst.

\vspace{1mm}

We stress also that the improvements in the SoD in Fig.~\ref{fig:sims} can be pursued even without accurate calibration of the $\bphi$ parameter since the constraints on $\fnl\bphi$ do not require priors on $\bphi$. The exercise on the right panel of Fig.~\ref{fig:sims} can be repeated for any observed galaxy sample by splitting it into two according to any measured galaxy property, and checking the improvements on the $\fnl\bphi$ constraints. Simulations can nonetheless still aid in this exploration through characterization of trends in $b_1$ and $\bphi$ as a function of galaxy selection criteria. For example, should some suite of different galaxy formation models consistently predict that splits by property $P$ yield larger values of $\bfac$, then this motivates splits based on $P$ to construct multi-tracer samples, even if the different models do not predict exactly the same values of $\bfac$. From Fig.~\ref{fig:sims} we identify from the IllustrisTNG model that splits by color or stellar mass are a promising strategy to improve the SoD of $\fnl$, but we caution that whether this holds also for other galaxy formation models is still unknown.

\section{The impact of priors on $\bphi$}
\label{sec:priors}

The discussion in the previous section shows that there is room to improve the probability to detect $\fnl\bphi$ that is not as affected by our ability to {\it precisely} predict the $\bphi$ parameter of real galaxies. However, ultimately we are interested in either the numerical value or upper bounds on $\fnl$ to draw more decisive conclusions about inflation. If $\bphi$ is exactly known, the ratio of the $\fnl$ precision for multi-tracer and single-tracer analyses $\sigma_{\fnl}^{\mathrm{MT}}/\sigma_{\fnl}^{\mathrm{ST}}$ is the inverse of the SoD ratio on the right of Fig.~\ref{fig:sims}. This is not the case in reality, however, where calibration methods for $\bphi$ (e.g.~based on simulations \cite{2010JCAP...07..013R, 2017MNRAS.468.3277B, 2020JCAP...12..013B, Voivodic:2020bec, 2022JCAP...01..033B, 2022JCAP...04..057B, 2023JCAP...01..023L}) are expected to come with some associated systematic offset and uncertainty.

In this section, we discuss the impact of priors on $\bphi$ in single-tracer and multi-tracer $\fnl$ constraints (see Refs.~\cite{2020JCAP...12..031B, 2021JCAP...05..015M, 2022JCAP...01..033B} for previous discussions for the single-tracer case). The results here are based on simulated likelihood analyses, i.e., we do not work with Fisher matrix forecasts to be able to quantify the impact of $\bphi$ misspecifications on both the mean and error bar of the $\fnl$ constraints. We keep assuming a fictitious survey with $V_S = 100\ {\rm Gpc^3}/h^3$ at $z=1$.

\subsection{The impact of systematic offsets in $\bphi$}

\begin{figure}
\centering
\includegraphics[width=.7\textwidth]{\pathtofigs 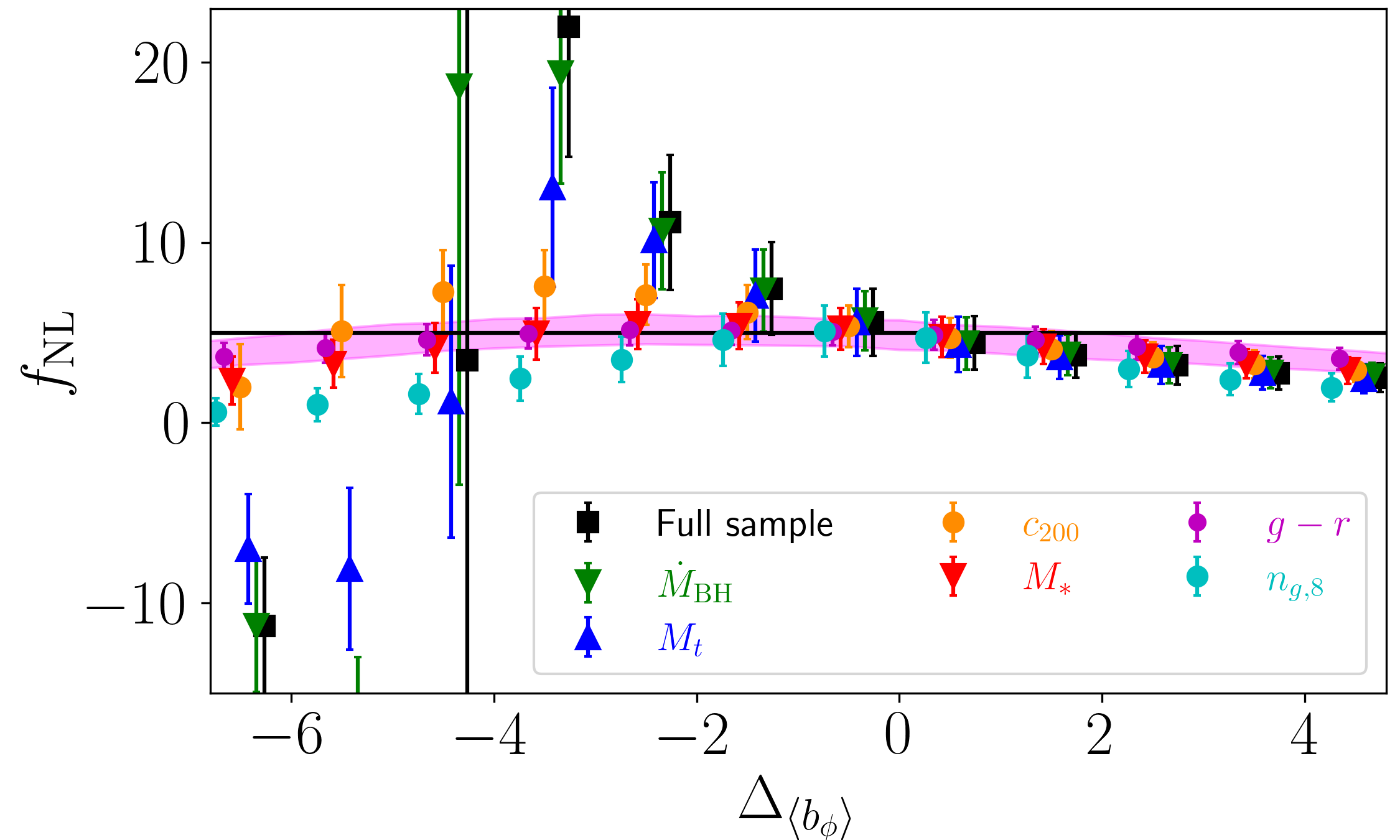}
\caption{Impact of a systematic offset $\Delta_{\bphia}$ between the assumed and true values of $\bphi$ on $\fnl$ constraints. The different symbols/colors are for the different samples we consider in this work, as labeled. The magenta band simply marks the $1\sigma$ uncertainty region for the $(g-r)$ multi-tracer split, which is the most robust to $\Delta_{\bphia}$. At a given $\Delta_{\bphia}$, the symbols are shifted horizontally for visualization; e.g., the third cluster of symbols from the left are all for  $\Delta_{\bphia} =-4.5$, which is the $\bphi$ zero-crossing of the full galaxy sample.}
\label{fig:offset}
\end{figure}

We start with the case where $\bphi$ is precisely but inaccurately known, i.e., we investigate the impact of a systematic offset $\Delta_{\bphia}$ defined as $\bphi^{A,B} = \langle\bphi\rangle^{A,B} + \Delta_{\bphia}$, where $\langle\bphi\rangle$ is the true value of the samples. We assume for simplicity the same offset $\Delta_{\bphia}$ for the sub-samples $A,B$ in multi-tracer and for the full sample in single-tracer. Figure \ref{fig:offset} shows the impact of $\Delta_{\bphia}$ on the $\fnl$ constraints for the simulation-inspired galaxy selections of Fig.~\ref{fig:sims}. The single-tracer result displays biased $\fnl$ constraints that follow the expectation $\fnl = \langle\fnl\rangle \langle\bphi\rangle / \left(\langle\bphi\rangle + \Delta_{\bphia}\right)$, where $\langle\fnl\rangle = 5$ is the true value. In particular, the constraints diverge when $\Delta_{\bphia} = \langle\bphi\rangle = -4.5$. The critical impact of systematic offsets like $\Delta_{\bphia}$ in single-tracer constraints using BOSS DR12 galaxy data has been recently investigated in Ref.~\cite{2022JCAP...11..013B}.

In contrast, the multi-tracer constraints appear more robust to $\Delta_{\bphia}$. While for $\Delta_{\bphia} > 0$ the impact is similar for multi-tracer and single-tracer, the multi-tracer constraints are visibly more robust to $\Delta_{\bphia} < 0$. The degree of robustness depends on the type of multi-tracer split, with the result in Fig.~\ref{fig:offset} displaying a rough correlation with the value of $\bfac$. The splits by $g-r$, $M_*$ and $c_{200}$ are the most robust, being effectively unbiased on the range $-6 \lesssim \Delta_{\bphia} \lesssim 1$. In contrast, the splits by $\dot{M}_{\rm BH}$ and $M_t$ perform closer to the single-tracer result.\footnote{
The log-likelihood from sample $A$ is $-2{\rm ln}\mathcal{L}^A = \left(\langle\fnl\bphi^A\rangle - \fnl\left(\langle\bphi^A\rangle + \Delta_{\bphia}\right)\right)^2/\sigma_A^2$ with $\sigma_{A} \equiv \sigma_{\fnl\bphi^A}$, and similarly for the other sample with $A \leftrightarrow B$. Ignoring the correlation between the two samples, the joint likelihood is the product $\mathcal{L}^A\mathcal{L}^B$, which has the following maximum likelihood estimator (the brackets $\langle\rangle$ indicate true values)
\bq\label{eq:mle}
\widehat{\fnl} = \langle{\fnl}\rangle\frac{\langle\bphi^A\rangle\left(\langle\bphi^A\rangle+\Delta_{\bphia}\right)/\sigma_A^2+\langle\bphi^B\rangle\left(\langle\bphi^B\rangle+\Delta_{\bphia}\right)/\sigma_B^2}{\left(\langle\bphi^A\rangle+\Delta_{\bphia}\right)^2/\sigma_A^2 + \left(\langle\bphi^B\rangle+\Delta_{\bphia}\right)^2/\sigma_B^2}\,.
\eq
This expression can be used to build some intuition about the robustness to $\Delta_{\bphia}$. For example, if $\bphi$ is misspecified as zero in single-tracer analyses, then this would imply no constraining power on $\fnl$. However, in multi-tracer analyses, even if $\Delta_{\bphia}$ leads to misspecifying one of the $\bphi$ as zero, one can still constrain $\fnl$ from the other sample (assuming $\bphi^A \neq \bphi^B$). Further, the robustness in Fig.~\ref{fig:offset} can be understood in terms of a {\it cancellation} of effects by the two samples $A, B$ when the assumed $\bphi^A$ and $\bphi^B$ bracket zero (i.e.~one is negative and the other positive), which for our samples happens when $\Delta_{\bphia} < 0$. More concretely, for some $b_\phi$ misspecifications the two samples shift $\fnl$ in opposite directions, yielding unbiased constraints. In contrast for single-tracer, any misspecification of $\bphi$ always leads to biased constraints.}

\subsection{Exploration of multi-tracer $\bphi$ prior requirements}

\begin{figure}
\centering
\includegraphics[width=1.0\textwidth]{\pathtofigs 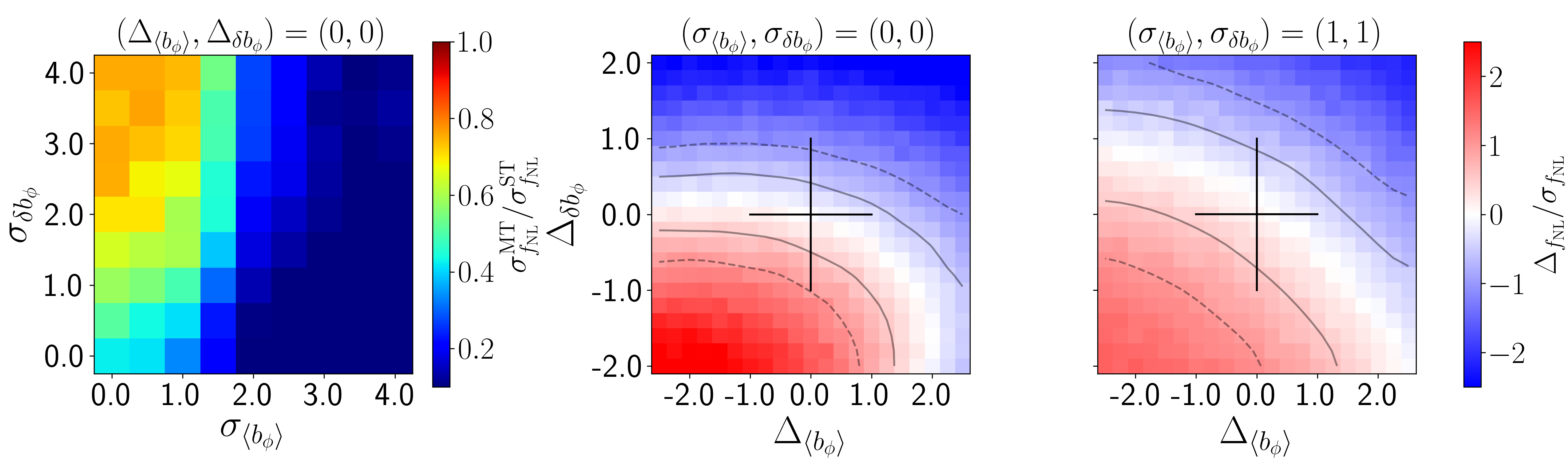}
\caption{Left: relative constraining power of multi- vs.~single-tracer analyses $\sigma^{\mathrm{MT}}_{\fnl}/\sigma^{\mathrm{ST}}_{\fnl}$ as a function of $\bphi$ uncertainties, in the absence of systematic offsets. Middle, right: relative bias $\Delta\fnl/\sigma_{f_\mathrm{NL}}$ as a function of systematic offsets in $\bphi$, for two different $\bphi$ uncertainties as labeled. The solid, dashed contours indicate $\Delta\fnl/\sigma_{\fnl}=0.5,1$, respectively. For guidance, the black cross marks unity deviations from the origin $(\Delta_{\bphia},\Delta_{\delta b_\phi})=(0,0)$. The results in this figure are for the $(g-r)$ multi-tracer case.}
\label{fig:prior}
\end{figure}

In order to more realistically relate multi-tracer $\bphi$ prior requirements to the single-tracer case, we consider now an extended parametrization of $\bphi$ uncertainties. We denote by $\bphi^{\rm ST}$ the value of the full sample, parametrize $\bphi$ for the two multi-tracer samples as
\bq
\bphi^{A} = \bphi^{\rm ST} + \delta\bphi,\ \ \ \ \bphi^{B} = \bphi^{\rm ST} - \delta\bphi,
\eq
and allow for systematic offsets and uncertainties on both $\bphi^{\rm ST}$ and $\delta\bphi$. This reflects the fact that astrophysical modeling uncertainties can manifest themselves in miscalibrations of the mean of the full sample, as well as the shifts in $\bphi$ caused by the multi-tracer selection criteria. Concretely, we assume a Gaussian prior for $\bphi^{\rm ST}$ with mean and standard deviation $\big(\mu, \sigma\big) = \big(\bphi^{\rm ST} + \Delta_{\bphia}, \sigma_{\bphia}\big)$, and the same for $\delta\bphi$ with $\big(\mu, \sigma\big) = \big(\delta\bphi + \Delta_{\delta\bphi}, \sigma_{\delta\bphi}\big)$. In this subsection we only show results for the $g-r$ multi-tracer split; we note that $\bphi$ prior requirements should be set relative to the underlying constraining power on $\fnl$, and so are by construction survey-, sample- and analysis-specific.

The left panel of Fig.~\ref{fig:prior} shows the impact on the $\fnl$ precision of the prior widths $\sigma_{\bphia}$ and $\sigma_{\delta\bphi}$; the colors indicate the ratio $\sigma_{\fnl}^{\rm MT}/\sigma_{\fnl}^{\rm ST}$ of the error bar in multi-tracer and single-tracer, and the result is at zero offset $\Delta_{\bphia} = \Delta_{\delta\bphi} = 0$. While the $\bphi$ uncertainty causes $\sigma_{\fnl}$ to increase for both multi- and single-tracer, the level of degradation of multi-tracer is kept significantly smaller. In this specific test, $\sigma_{\fnl}^{\rm ST}$ varies by a factor of $\sim 15$ across the plot, while $\sigma_{\fnl}^{\rm MT}$ increases by only $10\%$; this causes the strong left-to-right gradient.\footnote{Without prior information on $\bphi$, the likelihood in $\{\fnl, \bphi^A,\bphi^B\}$-space is constant along $\fnl\bphi = {\rm constant}$ surfaces. Thus, there is a special direction along which $\fnl$ can become arbitrarily large at fixed likelihood, and cannot therefore be constrained. The larger the number of $\fnl\bphi = {\rm constant}$ surfaces, the more delicate this balance is, and so the easier it is to break by priors. This explains the extra robustness of multi-tracer relative to single-tracer analyses.} Further, for our specific $g-r$ multi-tracer setup, the uncertainty on $\delta \bphi$ causes only minor degradation on the $\fnl$ precision (note the parameter $\sigma_{\delta\bphi}$ is only for the multi-tracer case). The apparent extra resilience to $\bphi$ uncertainties provides yet another motivation for multi-tracer $\fnl$ constraints. In this case, the left of Fig.~\ref{fig:prior} suggests that the $\bphi$ prior requirements may be less strict compared to single-tracer, which is opposite to the naive expectation that they would be stricter by the need to calibrate more than one $\bphi$ value.

The middle and right panels of Fig.~\ref{fig:prior} quantify relative biases $\Delta \fnl/\sigma_{f_\mathrm{NL}}$ in $\fnl$ estimates as a function of the systematic offset parameters $\Delta_{\bphia}$ and $\Delta_{\delta\bphi}$; the middle panel is for zero uncertainty $(\sigma_{\bphia},\sigma_{\delta b_\phi})=(0,0)$, and the right panel is for $(\sigma_{\bphia},\sigma_{\delta b_\phi})=(1,1)$. For orientation, the row at $\Delta_{\delta b_\phi}=0$ in the middle panel corresponds to the $g-r$ (magenta) symbols in Fig.~\ref{fig:offset}. The result shows that while certain extreme misspecifications of $\bphi$ can lead to $2\sigma$ biases (e.g.~$(\Delta_{\bphia}, \Delta_{\delta\bphi}) \approx  (-2.5, -2)$ in the middle panel), there is also a {\it order unity} region around the zero misspecification point (black cross) where the $\fnl$ biases remain small. As expected, the $\bphi$ prior requirements for $\Delta \fnl/\sigma_{f_\mathrm{NL}}$ are more relaxed for the case with $(\sigma_{\bphia},\sigma_{\delta b_\phi})=(1,1)$, mostly due to the degradation in constraining power. We stress again that our quantitative results are specific to our survey setup and simulation-inspired $g-r$ multi-tracer split. Figure \ref{fig:prior} serves nonetheless as an example for how to deduce $\bphi$ prior requirements for future real data analyses.

\section{Summary and Conclusions}
\label{sec:summary}

We revisited the potential of the multi-tracer technique to improve constraints on $\fnl$ using galaxy data. Using Fisher matrix and parameter sampling forecasts for an idealized survey, we (i) quantified the impact of galaxy selection criteria on the constraining power, and (ii) explored the robustness of multi- and single-tracer analyses to $\bphi$ misspecifications and uncertainties. Our main takeaways are:

\begin{itemize}

\item For multi-tracer analyses with two sub-samples $A, B$ the constraining power is $\propto \bfac$ \cite{2009PhRvL.102b1302S} (cf.~Fig.~\ref{fig:circle}). This is different, and leads in general to better constraints, than the naive expectation $\propto |b_1^A - b_1^B|$ based on universality-like relations $\bphi \propto b_1$.

\item Indeed, different equal-number multi-tracer splits in galaxy formation simulations can lead to markedly different $\bfac$. For IllustrisTNG galaxies, we find that splits based on $g-r$ color can more than double the significance of detection (SoD) of $\fnl\bphi \neq 0$ (cf.~Fig.~\ref{fig:sims}). This result follows explicitly from the fact that $g-r$ cuts do not satisfy universality-like relations.

\item Importantly, this ability to improve the SoD does not require priors on $\bphi$, and can even be pursed {\it blindly} by applying different multi-tracer splits according to any observed galaxy properties.

\item Compared to single-tracer, multi-tracer $\fnl$ constraints are also significantly more robust to $\bphi$ misspecifications (cf.~Fig.~\ref{fig:offset}) and $\bphi$ uncertainties (cf.~Fig.~\ref{fig:prior}). This provides extra incentive for multi-tracer analyses as a way to relax the requirements on $\bphi$ priors from numerical simulations.

\end{itemize}

As future work, it would be interesting to study multi-tracer splits based on other galaxy properties, as well as to investigate how different astrophysics models compare in their $b_1$, $\bphi$ predictions. This is important to help identify which multi-tracer splits are most promising, as well as to calibrate $\bphi$ priors. Our results motivate significantly investing in simulation campaigns towards improved chances to detect $\fnl \neq 0$ and robustly constrain its value using multi-tracer galaxy data analyses. 

\acknowledgments

We would like to thank Vincent Desjacques, Henry Gebhardt, Nick Kokron, Titouan Lazeyras and Fabian Schmidt for very useful comments and conversations. AB acknowledges support from the Excellence Cluster ORIGINS which is funded by the Deutsche Forschungsgemeinschaft (DFG, German Research Foundation) under Germany's Excellence Strategy - EXC-2094-390783311. EK is supported in part by an Alfred P. Sloan Research Fellowship, the David and Lucile Packard Foundation, and the SPHEREx project under a contract from the NASA/GODDARD Space Flight Center to the California Institute of Technology. Part of this work is based on High-Performance Computing resources supported by the University of Arizona TRIF, UITS, and Research, Innovation, and Impact (RII) and maintained by the UArizona Research Technologies department
\appendix
\section{Derivation of the Fisher information content in multi-tracer analyses}
\label{app:derivation}
For a galaxy sample split into two sub-samples $A$ and $B$, the Fisher information on $\fnl$ is defined as
\bq\label{eq:Finfo_app_1}
F_{\fnl}(k) &=& \frac{\partial {\bf D}^t(k)}{\partial \fnl} \cdot {\bf Cov}^{-1}(k) \cdot  \frac{\partial {\bf D}(k)}{\partial \fnl},
\eq
with the data vector given by
\bq\label{eq:D_app_1}
{\bf D}(k) &=& \Big\{P_{gg}^{A}(k), P_{gg}^{A\times B}(k), P_{gg}^{B}(k)\Big\},
\eq
and its covariance by (assuming only the contribution from the diagonal, disconnected piece)
\bq\label{eq:cov_app_1}
{\Cov}(k) = 2 \frac{(2\pi)^3}{V_sV_{k}}
\begin{pmatrix}
[P_{gg}^{A}(k)]^2 & P_{gg}^{A\times B}(k)P_{gg}^{A}(k) & [P_{gg}^{A\times B}(k)]^2 \\[1.5ex]
\cdots & \frac12\left[P_{gg}^{A}(k)P_{gg}^{B}(k) + [P_{gg}^{A\times B}(k)]^2\right] & P_{gg}^{B}(k)P_{gg}^{A\times B}(k) \\[1.5ex]
\cdots & \cdots & [P_{gg}^{B}(k)]^2
\end{pmatrix}
\,\,.
\eq
Rather than directly evaluating Eq.~(\ref{eq:Finfo_app_1}), we perform a few intermediate derivation steps as in Ref.~\cite{2009PhRvL.102b1302S} to help to highlight the origin of the power of the multi-tracer approach. The power spectra in Eq.~(\ref{eq:D_app_1}) can be expressed as 
\bq\label{eq:spectra_app_1}
P_{gg}^{A}(k,z) &=& R(k)^2 f^B(k)^2 P_{mm}(k) + P_{\eps}^A, \nonumber \\
P_{gg}^{A\times B}(k,z) &=& R(k) f^B(k)^2 P_{mm}(k), \nonumber \\
P_{gg}^{B}(k,z) &=& f^B(k)^2 P_{mm}(k) + P_{\eps}^B, 
\eq
where 
\bq\label{eq:spectra_app_2}
f^A(k) &=& b_1^A + \fnl\bphi^A\minv, \nonumber \\
f^B(k) &=& b_1^B + \fnl\bphi^B\minv, \nonumber \\
R(k)   &=& \frac{f^A(k)}{f^B(k)}.
\eq
Note that the quantity $R(k)$ measures the ratio of the deterministic (i.e.~cosmological, or non-stochastic) parts of the galaxy power spectrum. As pointed out in the original multi-tracer discussion of Ref.~\cite{2009PhRvL.102b1302S}, the power of the multi-tracer formalism relies in the observation that in the cosmic variance limit (small $P_{\eps}$) the quantity $R$ can be measured with infinite precision, and thus so can in principle any other parameter that $R$ depends on, including $\fnl$. Under the parametrization of Eq.~(\ref{eq:spectra_app_1}) the data vector depends only on the variables $R$ and $f^B$, and the Fisher information can be expressed as
\bq\label{eq:Finfo_app_2}
F_{\fnl} &=& \sum_{\lambda, \lambda' = R, f^B} \frac{\partial {\bf \lambda}}{\partial \fnl} F_{\lambda\lambda'} \frac{\partial {\bf \lambda'}}{\partial \fnl} \nonumber \\
&=& F_R \left(\frac{\partial R}{\partial \fnl}\right)^2 + F_{f^B} \left(\frac{\partial f^B}{\partial \fnl}\right)^2 + 2F_{R,f^B} \left(\frac{\partial R}{\partial \fnl}\right) \left(\frac{\partial f^B}{\partial \fnl}\right),
\eq
where $F_R$, $F_{f^B}$ and $F_{F, f^B}$ are, respectively, the Fisher information on $R$, $f^B$ and their cross term. These are given by
\bq\label{eq:Finfo_app_3}
F_R &=&  2\frac{V_S V_k}{2(2\pi)^3} \frac{X^A+X^B + X^AX^B + 2(X^B)^2}{R^2(X^A+X^B + X^AX^B)^2} \nonumber \\
F_{f_B} &=&  4\frac{V_S V_k}{2(2\pi)^3} \frac{(X^A+X^B)^2}{(f^B)^2(X^A+X^B + X^AX^B)^2} \nonumber \\
F_{R,f_B} &=&  4\frac{V_S V_k}{2(2\pi)^3} \frac{X^B(X^A+X^B)}{Rf^B(X^A+X^B + X^AX^B)^2} \nonumber \\
\eq
The key observation is that, in the cosmic variance limit $X_A, X_B\to 0$, the Fisher information on $F_R$ diverges as 
\bq\label{eq:Finfo_app_4}
F_R \stackrel{}{\longrightarrow} 2\frac{V_S V_k}{2(2\pi)^3} \frac{1}{R^2\left(X^B + X^A\right)}, 
\eq
while $F_{f^B}$ and $F_{R, f^B}$ asymptote to a constant. In this limit the power spectra of samples $A$ and $B$ is effectively deterministic and their ratio can be measured perfectly as both samples formed out of the same realization of the initial density field. From Eq.~(\ref{eq:Finfo_app_2}) we thus have in this limit that
\bq\label{eq:Finfo_app_5}
F_{\fnl} \stackrel{}{\longrightarrow}  2\frac{V_S V_k}{2(2\pi)^3} \frac{1}{X^B + X^A} \left(\frac{\left(b_1^B \bphi^A - b_1^A\bphi^B\right)\minv}{R(f^B)^2}\right)^2 + \mathcal{O}\left(X^A, X^B\right),
\eq
which matches Eq.~(\ref{eq:fisher_info_mt_2}). As discussed in the main body of the paper, this equation tells us that, in the cosmic variance limit, the information content on $\fnl$ increases with $\left|b_1^B \bphi^A - b_1^A\bphi^B\right|$, which is why multi-tracer splits that maximize this quantity can return better constraints on $\fnl$. 

\bibliographystyle{utphys}
\bibliography{REFS}

\pagebreak\end{document}